\documentclass[aps]{revtex4}%
\usepackage{amsfonts}
\usepackage{amsmath}
\usepackage{amssymb}
\usepackage{graphicx}
\usepackage{hyperref}%
\providecommand{\U}[1]{\protect\rule{.1in}{.1in}}

\begin{document}
\title{Comment on "New apparatus design for high-precision measurement of G with atom interferometry"}
\author{B. Dubetsky}
\affiliation{Independent researcher, 1849 S Ocean Dr, Apt 207, Hallandale, FL 33009, Usa}
\date{\today}
\begin{abstract}
It is shown that even in the case of a negligibly small change in the gradient
of the gravitational field of the mass source in the axial direction, the
dependence of this gradient in the radial direction leads to a systematic
error in Newton gravitational constant value. The magnitude of this error is
calculated for two configurations of the field. For both configurations it was
found that this error is larger than the inaccuracy predicted in the article
by M. Jain et al [Eur. Phys. J. D \textbf{75}, 197 (2021)]. In addition, we
found the geometry of the source mass, for which this systematic error disappears.

\end{abstract}
\maketitle
\preprint{HEP/123-qed}
%

\twocolumngrid

Atom interference \cite{c1} can be used to measure Newtonian gravitational
constant $G$ \cite{c2}. The best measurement of this constant was performed in
the experiment \cite{c3}. It was stated that the measurement accuracy of 148
ppm was achieved. However, we have attempted \cite{c4} to build an error
budget for the conditions of the experiment \cite{c3}, we have shown that due
to uncertainties in the initial position of the centers of atomic clouds in
the coordinate and velocity spaces, the accuracy of the measurement of $G$ is
not less than $275$ppm. We also found that the nonlinear dependence of the
phase of atomic interferometers (AIs) on these uncertainties should have led
to a systematic error of $-199$ppm in the measurement of $G$

An alternative measurement method is proposed in the article \cite{c5}. If the
gravitational field of the source mass has a strictly linear dependence on the
coordinates along the atomic trajectory, then with a certain change in the
effective wave vector of the second Raman pulse, the dependence of the phase
AI on the initial coordinates and velocities disappears \cite{c6}. This effect
was observed in the articles \cite{c5,c7}. In the articles \cite{c8,c9}, the
budget of errors of the alternative method was studied. Two configurations of
the source mass were considered in the article \cite{c9}. Based on the built
error budget, the authors came to the conclusion that the accuracy of the
measurement of $G$ can be $13$ppm and 5ppm for configurations I and II,
respectively. In this commentary, we will calculate the contributions to the
relative standard deviation (RSD) and systematic error (RSE) due,
respectively, to uncertainties in the radial position of the center of the
atomic cloud and the finite radial size of the atomic cloud. We will show that
the RSEs will exceed the measurement accuracy predicted in \cite{c9}.

Near the center , the axial component of the gravitational field in the
presence of a source mass is given by%

\begin{subequations}
\label{1}%
\begin{align}
g_{z}\left(  \mathbf{x}\right)   &  =g+\gamma\left(  r\right)  z,\label{1a}\\
\gamma\left(  r\right)   &  =\gamma\left(  1+\varepsilon r^{2}\right)
,\label{1b}\\
\gamma &  =\Gamma_{33}^{\left(  1\right)  },\label{1c}\\
\varepsilon &  =\Gamma_{3311}^{\left(  3\right)  }/2\gamma, \label{1d}%
\end{align}
where $g$ is the gravitational field of the Earth, $\mathbf{x}=\left(
\mathbf{x}_{\perp},z\right)  ,$ $\mathbf{x}_{\perp}=\left(  x,y\right)  $ is
the radial component of the vector $\mathbf{x},$ $r=\left\vert \mathbf{x}%
_{\perp}\right\vert $,$\Gamma^{\left(  i\right)  }$ is the order $i$
gravity-gradient tensor of the source mass gravitational field \cite{c10}.
Suppose that the basic idea of the alternative method \cite{c5,c8,c9} is
implemented, one can neglect completely nonlinear terms in the field
dependence on the axial coordinate $z$. Let's also assume that one can neglect
the recoil effect, that the atoms are launched in a strictly axial direction,
their initial velocity $\mathbf{v}\left(  0\right)  =\left(  0,0,v_{0}\right)
$, and that there is no radial component of the gravitational field. Then,
when an atom interacts with 3 Roman pulses, the phase of the atomic
interferometer is given by \cite{c11}%
\end{subequations}
\begin{equation}
\phi=k_{1}z\left(  T_{1}\right)  -2k_{2}z\left(  T_{2}\right)  +k_{3}z\left(
T_{3}\right)  , \label{2}%
\end{equation}
where $k_{m}$ is the effective wave vector of the Raman pulse $m$,
$T_{m}=T_{1}+\left(  m-1\right)  T,$ $T_{1}$ is the time delay between first
pulse and the atoms launching moment, $T$ is the interrogation time, $z\left(
t\right)  $ is the axial component of the atomic trajectory. Considering the
second term in Eq. (\ref{1a}) as a small addition, one gets%

\begin{equation}
z\left(  t\right)  =z_{0}+v_{0}t+\dfrac{1}{2}gt^{2}+\gamma\left(  r\right)
\left(  \dfrac{1}{2}z_{0}t^{2}+\dfrac{1}{6}v_{0}t^{3}+\dfrac{1}{24}%
gt^{4}\right)  . \label{3}%
\end{equation}
In the absence of radial components of the gravitational field and the initial
velocity of the atom, $r$ is the distance from the axis of the source mass to
the position of the atom. For the choice of wave vectors $k_{1}=k_{3}%
=k,k_{2}=k+\Delta k_{2}$ proposed in \cite{c6}, one has \cite{c12}%
\begin{align}
\phi &  =kgT^{2}+\left[  z_{0}+v_{0}\left(  T_{1}+T\right)  \right]  \left[
k\gamma\left(  r\right)  T^{2}-2\Delta k_{2}\right] \nonumber\\
&  +g\left[  k\gamma\left(  r\right)  \dfrac{6T_{1}^{2}T^{2}+12T_{1}%
T^{3}+7T^{4}}{12}\right. \nonumber\\
&  \left.  -\Delta k_{2}\left(  T_{1}^{2}+2TT_{1}+T^{2}\right)  \right]  .
\label{4}%
\end{align}

One sees that at
\begin{equation}
\gamma\left(  r\right)  =\dfrac{2\Delta k_{2}}{kT^{2}}, \label{5}%
\end{equation}
the AI phase does not depend on the atomic axial coordinates and velocity
$\left(  z_{0},v_{0}\right)  $ \cite{c6}, which makes it possible to measure
the gradient of the gravitational field of the source mass $\gamma$ and
Newtonian gravitational constant $G$ \cite{c5,c8,c9}. The dependence of
$\gamma$ on the radial atomic coordinates $\mathbf{x}_{\perp}$ leads to the
RSE $s$ and RSD $\sigma$ of this measurement.

For differential technique, one uses two AIs. Let's assume that the atomic
clouds in these interferometers are identical. Then, averaging (\ref{5}) over
the atomic cloud, one finds%

\begin{equation}
s=-\varepsilon\left(  a_{x}^{2}+a_{y}^{2}\right)  , \label{6}%
\end{equation}
where $\left(  a_{x},a_{y}\right)  =\left[  \int d\mathbf{x}_{\perp}\left(
x^{2},y^{2}\right)  f\left(  \mathbf{x}_{\perp}\right)  \right]  ^{1/2}$ and
$f\left(  \mathbf{x}_{\perp}\right)  $ are the radii of the atomic cloud and
the radial atomic distribution function.

Uncertainties in the position of cloud centers with standard deviations
$\left(  \sigma_{x},\sigma_{y}\right)  $ lead to uncertainty $\gamma$ with RSD
$\sigma_{\gamma}$. The usual formula for RSD
\begin{equation}
\sigma_{\gamma}=\gamma^{-1}\left[  \left(  \partial\gamma/\partial x\right)
_{\mathbf{x}_{\perp}=0}^{2}\sigma_{x}^{2}+\left(  \partial\gamma/\partial
y\right)  _{\mathbf{x}_{\perp}=0}^{2}\sigma_{y}^{2}\right]  ^{1/2} \label{7}%
\end{equation}
leads to a zero result, because due to the symmetry of the source mass, its
center is an extreme point, $\left(  \partial\gamma/\partial\mathbf{x}_{\perp
}\right)  _{\mathbf{x}_{\perp}=0}=0$. In the article \cite{c4}, we obtained
the formula (16a) for RSD, which one can use at extreme points. From it, one
will get that
\begin{align}
\sigma_{\gamma}  &  =\gamma^{-1}2^{-1/2}\left[  \left(  \partial^{2}%
\gamma/\partial x^{2}\right)  _{\mathbf{x}_{\perp}=0}^{2}\sigma_{x}%
^{4}+\left(  \partial^{2}\gamma/\partial y^{2}\right)  _{\mathbf{x}_{\perp}%
=0}^{2}\sigma_{y}^{4}\right]  ^{1/2}\nonumber\\
&  =\left\vert \varepsilon\right\vert \sqrt{2\left(  \sigma_{x}^{4}+\sigma
_{y}^{4}\right)  }. \label{8}%
\end{align}
Here we assumed that the cummulants do not contribute, $\kappa\left(
x\right)  =\kappa\left(  y\right)  =0$. Quadratic terms in the dependence of
signal variations on the uncertainties of atomic variables lead to an
additional RSE $s^{\prime},$ for which from the Eq. (16b) in \cite{c4} one
gets
\begin{equation}
s^{\prime}=-\varepsilon\left(  \sigma_{x}^{2}+\sigma_{y}^{2}\right)  .
\label{9}%
\end{equation}

We applied the results (\ref{6}, \ref{8}, \ref{9}) for the source mass
configurations proposed in \cite{c9}. For the configuration I, using the
formula for the axial component of the gravitational field of the cylinder,
derived in \cite{c4}, one obtains $\varepsilon=-1.45$m$^{-2}$. From the Table
2 in the Ref. \cite{c9} one can conclude that the radius of the atomic cloud
$a=\sqrt{a_{x}^{2}+a_{y}^{2}}$ is in the range of $2.6$mm$<a<3.4$mm. Then for
RSE (\ref{6}) one obtains
\begin{equation}
9.8\text{ppm }<s<17\text{ppm.} \label{10}%
\end{equation}
This RSE may be greater than the measurement accuracy of 13ppm predicted in
\cite{c9}. From the same Table, one can conclude that the uncertainties in the
position of the atomic cloud $\sigma_{x}=\sigma_{y}=1$mm. Then $\sigma
_{\gamma}=s^{\prime}=2.9$ppm.

For the configuration II, one finds that
\begin{equation}
\varepsilon=\dfrac{3}{4}\left[  \dfrac{y}{\left(  h^{2}/4+y\right)  ^{5/2}%
}\right]  _{y=R_{1}^{2}}^{y=R_{2}^{2}}\left/  \left[  \dfrac{1}{\sqrt
{h^{2}/4+y}}\right]  _{y=R_{1}^{2}}^{y=R_{2}^{2}}\right.  , \label{11}%
\end{equation}
where $R_{1},R_{2},$ and $h$ are the inner and outer radii, and the height of
the source mass. At $\left\{  R_{1},R_{2},h\right\}  =\left\{  0.15\text{m,
}0.52\text{m, }0.84\text{m}\right\}  $ $\varepsilon=-0.755$m$^{-2}$, and at
the same values of the radial radii of the atomic cloud, one gets for the RSE
$s$
\begin{equation}
5.1\text{ppm }<s<8.7\text{ppm,} \label{12}%
\end{equation}
which, again, is more than the accuracy of measuring $G$ 5ppm predicted in
\cite{c9}. At the same time, RSD (\ref{8}) and RSE (\ref{9}) are equal to
$\sigma_{\gamma}=s^{\prime}=0.75$ppm.

\begin{figure}[ptb]%
\centering
\includegraphics[
height=3.495in,
width=3.3878in
]%
{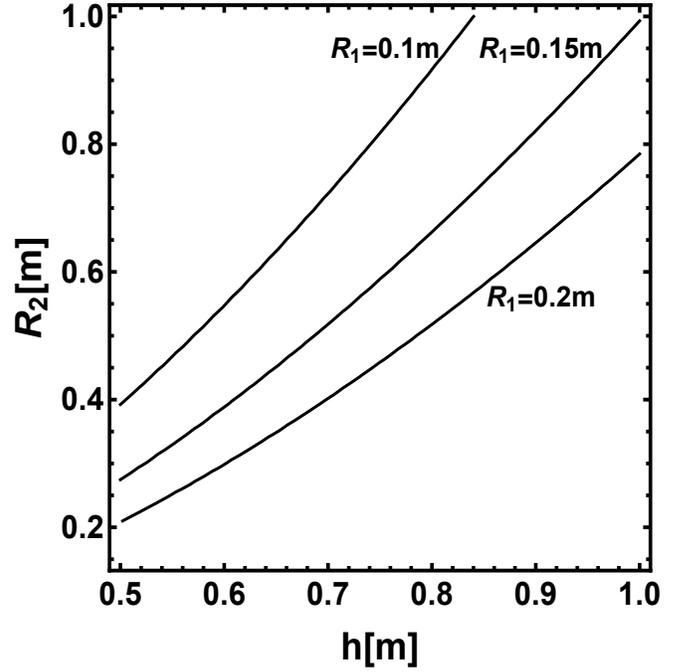}%
\caption{Contour plots of the solution of the equation $\varepsilon=0$ in the
plane of the height and the outer radius of the hollow cylinder $\left\{
h,R_{2}\right\}  $ at the different values of the inner radius $R_{1}$ }%
\label{f1}%
\end{figure}

We propose in this comment to choose other sizes of the source mass at which
the parameter (\ref{11}) disappears. The numerical solution of the equation
$\varepsilon=0$ is shown in the figure \ref{f1}.
So, for example, with the same internal and external radius of a hollow
cylinder, $\left\{  R_{1},R_{2}\right\}  =\left\{  0.15\text{m, }%
0.52\text{m}\right\}  $, it is sufficient to reduce the height of the cylinder
to the value $h=0.701$m to turn off the systematic and statistical measurement
errors caused by the finite radial size of the atomic cloud.

Conclusion. The finite size of the atomic cloud leads to a systematic error
that is greater than the accuracy of measuring Newtonian gravitational
constant $G$, predicted in the article Ref. \cite{c9}


\begin{thebibliography}{99}                                                                                               %


\bibitem {c1}B Ya Dubetskii, A. P. Kazantsev, V. P. Chebotayev, V. P.
Yakovlev, Interference of atoms and formation of atomic spatial arrays in
light fields, \textit{Pis'ma Zh. Eksp. Teor. Fiz.} \textbf{39}, 531 (1984)
[\textit{JETP Lett}. \textbf{39}, 649 (1984)].

\bibitem {c2}J. B. Fixler, G. T. Foster, J. M. McGuirkand, M. A. Kasevich,
Atom Interferometer Measurement of the Newtonian Constant of Gravity,
\textit{Science}
\textbf{\href{https://doi.org/10.1126/science.1135459}{\textbf{315},
74\textbf{.}(2007).}}

\bibitem {c3}G. Rosi, F. Sorrentino, L. Cacciapuoti, M. Prevedelli \& G. M.
Tino, Precision measurement of the Newtonian gravitational constant using cold
atoms, \textit{Nature}
\textbf{\href{https://doi.org/10.1038/nature13433}{\textbf{510}, 518 (2014).}}

\bibitem {c4}B Dubetsky, Newtonian gravitational constant measurement. All
atomic variables become extreme when using a source mass consisting of three
or more parts, Metrologia
\textbf{\href{https://doi.org/10.1088/1681-7575/abc92e}{\textbf{58}, 015004
(2021).}}

\bibitem {c5}G. D'Amico, G. Rosi, S. Zhan, L. Cacciapuoti, M. Fattori, and G.
M. Tino, Canceling the Gravity Gradient Phase Shift in Atom Interferometry,
Phys. Rev. Lett.
\textbf{\href{https://doi.org/10.1103/PhysRevLett.119.253201}{\textbf{119},
253201 (2017).}}

\bibitem {c6}A. Roura, Circumventing Heisenberg's Uncertainty Principle in
Atom Interferometry Tests of the Equivalence Principle, Phys. Rev. Lett.
\textbf{\href{https://doi.org/10.1103/PhysRevLett.118.160401}{\textbf{118},
160401 (2017).}}

\bibitem {c7}C. Overstreet, P. Asenbaum, T. Kovachy, R. Notermans, J. M.
Hogan, M. A. Kasevich, Effective inertial frame in an atom interferometric
test of the equivalence principle, Phys. Rev. Lett.
\textbf{\href{https://doi.org/10.1103/PhysRevLett.120.183604}{\textbf{120},
183604 (2018).}}

\bibitem {c8}G. Rosi, A proposed atom interferometry determination of $G$ at
$10^{-5}$ using a cold atomic fountain, Metrologia
\textbf{\href{https://iopscience.iop.org/article/10.1088/1681-7575/aa8fd8}{\textbf{55}%
, 50 (2017).}}

\bibitem {c9}M. Jain, G. M. Tino, L. Cacciapuoti, and G. Rosi, New apparatus
design for high-precision measurement of G with atom interferometry, Eur.
Phys. J. D
\textbf{\href{https://doi.org/10.1140/epjd/s10053-021-00212-6}{\textbf{75},197
(2021).}}

\bibitem {c10}In the Eq. (\ref{1a}) we neglected the gravity-gradient tensor
of the Earth's field. It can be excluded, when the double differential
technique \cite{c2} is used. We also took into account that the third order
tensors $\Gamma_{3311}^{\left(  3\right)  }$ and $\Gamma_{3322}^{\left(
3\right)  }$ coincide for both source mass configurations proposed in
\cite{c9}.

\bibitem {c11}B. Dubetsky, Asymmetric Mach-Zehnder atom interferometers,
\href{https://arxiv.org/abs/1710.00020v6}{arXiv:1710.00020v6
[physics.atom-ph].}

\bibitem {c12}B. Dubetsky, Comment on \textquotedblleft Circumventing
Heisenberg's Uncertainty Principle in Atom Interferometry Tests of the
Equivalence Principle,
\href{https://doi.org/10.1103/PhysRevLett.121.128903}{Phys. Rev. Lett,
121.128903 (2018).}
\end{thebibliography}
\end{document}